\documentclass[aps,prl,twocolumn,showpacs]{revtex4}
\usepackage{mathrsfs}
\usepackage{graphicx}
\newcommand{\D}{\displaystyle}
 \newcommand{\beq}{\begin{eqnarray}}
 \newcommand{\eeq}{\end{eqnarray}}
 \newcommand{\V}{ \mathscr{V} }

\newcommand{\Gam}{{\mathit \Gamma}}
\newcommand{\Om}{{\mathit \Omega}}

\newcommand{\BPsi}{ \mbox{\boldmath $\Psi$} }
\renewcommand{\d}{{\rm d}}
\begin{document}

\title{Perfect Reflection of Light by an Oscillating Dipole}

\author {G. Zumofen, N.M. Mojarad,  V. Sandoghdar, and  M. Agio}

\affiliation{Nano-Optics Group, Laboratory of Physical Chemistry, ETH Zurich, CH-8093 Zurich, Switzerland}

\date{\today}

\begin{abstract}

We show theoretically that a directional dipole wave can be
perfectly reflected by a single point-like oscillating dipole.
Furthermore, we find that in the case of a strongly focused plane
wave up to  85\,\% of the incident light can be reflected by
the dipole. Our results hold for the full spectrum of the
electromagnetic interactions and have immediate implications for
achieving strong coupling between a single propagating photon and a
single quantum emitter.

\end{abstract}


\maketitle

Common treatments of light-matter interaction consider the
excitation light to consist of a homogeneous field of area $\cal A$,
and often use the concept of a cross section $\sigma$ to arrive at
the probability $\sigma/\cal A$ for exciting an atom. In
conventional spectroscopy experiments, this ratio is very small
because either $\sigma$ is reduced by various broadening effects, or
$\cal A$ is large for technical reasons. However, recent experiments
have shown that it is possible to overcome these difficulties for
the optical excitation of single molecules, quantum dots, or atoms
~\cite{GWB,WGH,GSD,VAD,TCA}. The intriguing question that arises is
whether the experimentally observed coupling efficiencies are close
or far from theoretical limits. In particular, is it possible to
excite an atom with probability equal to one by a single
photon~\cite{SMK}?  Is it possible for an atom to imprint a large
phase shift on a photon that passes by?

On the theoretical side, the interaction of freely propagating
photons with the dipolar transition of a two-level system (TLS) has
been addressed for a quasi-one-dimensional case with emphasis on the
quantum statistics of the incident light~\cite{KC}. In three
dimensions, methods of expansion of the focused beam in terms of
vectorial mode functions and decomposition of the focused beam in
dipolar and non-dipolar vectorial modes have been
employed~\cite{EK,Enk}. These latter studies concluded that only the
dipolar component of the excitation light can couple to a dipole and
that the transmitted power is only weakly attenuated.
A more recent work finds that it is possible to absorb a photon with
near to one probability in a three-level system~\cite{PI}.
Interestingly
and somewhat in parallel, the literature on the interaction of a TLS
with light confined to a waveguide claims that very strong
attenuation is possible~\cite{DHR,CSD,SF}. In this Letter, we
examine the interaction of different light fields with a dipolar
emitter in the framework of Debye diffraction and vectorial
multipole expansion. We show that an incident directional dipolar
wave experiences strong coupling to the emitter and is fully
reflected. We first consider a classical oscillator and then extend
the analysis to a TLS.

\begin{figure}
\centerline{\includegraphics[width=0.95\columnwidth]{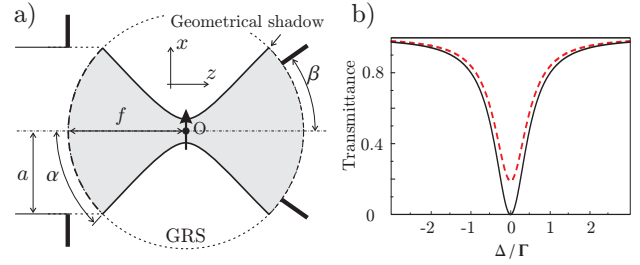}}
\caption{ a) Incident light propagating along the
$z$-axis is
focused with a spherical phase front onto a
dipole placed in vacuum. GRS: Gaussian reference sphere, a: entrance-aperture radius,
$\alpha$: entrance half angle, $\beta$: collection half angle, f: focal length.
b) The dashed curve plots the transmittance  for the
focused plane wave and displays an attenuation of ${\cal T} \simeq$ 80~\% at
resonance for $\alpha=\beta=\pi/3$. The solid curve shows that a
directional dipolar wave can be completely attenuated for $\alpha=\beta=\pi/2$. }
\end{figure}

The classical interaction of light with an oscillating point-like
dipole located at the origin $\bf \rm O$ is described by the
Abraham-Lorentz equation~\cite{Jac}. After calculating the
differential scattering cross section, one can
arrive at the total scattered power~\cite{Jac,Sup}
\beq
P_{\rm sca}= \frac 12 c \epsilon_0 \int_{4 \pi} r^2
        \left| {\bf E}_{\rm sca}({\bf r}) \right|^2 \d \Om
        =2 \, c \, {W}_{\rm inc}^{\rm el} ({\rm O}) \sigma ~,
\label{eq-Pscat}
\eeq
where ${\bf E}_{\rm sca}$ is the field
scattered by the dipole, and the distance $r$ lies in the far field,
$kr \gg 1$. ${W}_{\rm inc}^{\rm el}({\rm O}) = \epsilon_0 |{\bf E_
{\rm inc}}({\rm O})|^2/4 $ is the time-averaged electric energy
density at $\bf \rm O$. The parameter
\beq
\sigma = \sigma_0 \frac{
\Gam^2 }{ 4 \Delta^2 + \Gam^2}~,
\label{eq-sigma}
\eeq
denotes the
total scattering cross section of the oscillator where $\Gam$ is the
damping rate dictated by radiation reaction, and $\Delta =
\omega_{\rm L} - \omega_0$ is the detuning between the incident
light and oscillator frequencies $\omega_{\rm L}$ and $\omega_0$,
respectively. The quantity $\sigma_0 =  3 \lambda^2 /( 2 \pi )$
denotes the cross section on resonance ( $\Delta=0$). We now
consider the scattering ratio~\cite{Sup},
\beq
  {\cal K}_0= \frac{P_{\rm sca}}{P_{\rm inc}}
  =   \frac{ 2 \, c \, {W}_{\rm inc}^{\rm el} ({\rm O}) \sigma_0 }
    { \int   {\bf S}({\bf r}) \cdot {\bf n} ~ \d^2 r  }
    =   \frac{ \sigma_0 }
    { \cal A  } ~,
\label{eq-K}
\eeq
at resonance. Here $P_{\rm inc}$ is the incident
power, ${\bf S}$ is the time-averaged Poynting vector of the
incident field, and $\bf n$ is a unit vector normal to the
integration surface. The integration can be taken over a plane at
the incident aperture, over the Gaussian reference sphere (GRS), or
over the focal plane.

The derivation of Eqs.~(\ref{eq-Pscat}) and (\ref{eq-K}) is based on
the fact that the oscillator interacts only with the electric field
at the location of the oscillator, irrespective of whether the field
is homogeneous as for a plane wave, or inhomogeneous as in the focal
region of a strongly focused beam~\cite{RW}. Thus, $\sigma$ can be
treated as a universal quantity for a point-like oscillator
regardless of the modal properties of the excitation light. The
quantity $\cal A$ introduced in Eq.~(\ref{eq-Pscat}) represents an
effective focal area and depends implicity on $\lambda$ through the
diffraction phenomenon. It is closely related to the normalized
energy density ${W}_{\rm inc}^{\rm el}/P_{\rm inc}$, which has been
studied for various focal systems \cite{SL,ST,She}.  The peculiarities of
the incident field enter ${\cal K}_0$ via $\cal A$. Consequently, as
we will show below, the problem of minimal transmittance is shifted
to that of a minimal $\cal A$, and a strong photon-oscillator
interaction is reachable for ${\cal K}_0 \gtrsim 1$ \cite{DHR}.

We first consider an incident $x$-polarized plane wave of amplitude $E_0$. The
integration in Eq.~(\ref{eq-K}) over the incident aperture is
straightforward and yields $P_{\rm inc} = \frac 12  c \epsilon_0
E_0^2 \pi a^2$~\cite{Sup}, where $a$ is the radius of the entrance aperture.
We also have ${W}_{\rm inc}^{\rm el}({\rm
O})=\epsilon_0 (\pi f E_{\rm 0} \left| {\cal I}_0({\rm O}) \right|/2 \lambda)^2$
where $f$ is the focal length of the focusing system and ${\cal
I}_0({\rm O})$ is a diffraction integral~\cite{RW,Sup}. The
resulting value of ${\cal A}$ then yields
\beq
 {\cal K}_0 =\frac{128}{75} \frac 1 {\sin^2 \alpha}\left( 1  - \frac 18   ( 5 +
   3 \cos \alpha )\cos^{3/2} \alpha \right)^2~,
\label{eq-K2}
\eeq
where $\alpha$ specifies the incident solid angle
$\Om_\alpha$ (see Fig.~1a). For $\alpha=\pi/2$, ${\cal K}_0$ reaches
the maximum value of $128/75 \simeq 1.7$.
 Assuming a backward and forward half space
and accounting that half of the power is scattered in each direction,
it follows that up to 85\,\% of the incident light is
reflected into the backward half space. For this configuration
the reflectance and transmittance are thus limited
to ${\cal R} \stackrel{<}{\sim} 0.85$ and
${\cal T} = 1 - {\cal R} \gtrsim 0.15$, respectively.

An alternative way of performing the integration in Eq.~(\ref{eq-K})
is to consider the FP. Because the intensity has cylindrical
symmetry about the optical axis, the electric and the magnetic
energy densities are equal at the focal spot~\cite{RW} so that $2
c {W}_{\rm inc}^{\rm el}({\rm O}) = S_z({\rm O})$. The calculation
of $\cal A$ then becomes \beq
 {\cal A} =  \frac{ \int_{\rm FP} {S}_z \, \d^2 r }{ {S}_z({\rm O})}
  = \frac{   \int_{\rm FP} \big(  \left| {\cal I}_0 \right|^2
     - \left| {\cal I}_2 \right|^2 \big) \d^2 r  }
     { \left| {\cal I}_0({\rm O})  \right|^2}~,
\label{eq-A}
\eeq
where ${\cal I}_2$ is again a diffraction
integral~\cite{RW}. The integration in the numerator turns out to
be straightforward when an orthogonality relationship for Bessel
functions is considered~\cite{Sup}. We note that the fields in a
strongly focused beam show vortices in the FP~\cite{Sta} so that
$S_z$ takes on positive and negative values as shown in
Fig.~2a~\cite{RW}. Thus, in general ${S}_z(\textbf{r})$ cannot be
substituted by $2 c {W}_{\rm inc}^{\rm el}(\textbf{r})$, which is
a positive quantity. We remark in passing that Ref.~\cite{EK}
predicts a much lower value than $1.7$ for a quantity equivalent to
our parameter ${\cal K}_0$. We believe one of the origins of this
discrepancy is that Ref.~\cite{EK} takes the integrand in the
definition of $\cal A$ to be ${W}_{\rm inc}^{\rm el}(\textbf{r})$.

In order to derive an upper limit of ${\cal K}_0$ for the general
class of transverse axially symmetric systems, we consider the field
produced by the combination of an electric and a magnetic dipole which
has been suggested for optimal focusing~\cite{DS,SL}. To emulate
such a field, one considers the emission field patterns at the GRS
of virtual electric and magnetic dipoles orthogonal to each other
and placed at O and then reverses the field propagation. Using Eq.
(\ref{eq-K}) for the calculation of $\cal A$ we obtain~\cite{Sup}
\beq
 {\cal K}_0= \frac 14 \left( 7 - 3 \cos \alpha - 3 \cos^2 \alpha - \cos^3 \alpha \right)~.
\label{eq-p+m} \eeq At $\alpha = \pi/2$,  ${\cal K}_0=7/4$
establishes the maximum value for transverse axially symmetric
systems. This is only slightly larger than 128/75 obtained for the
plane wave.

We next abandon the restriction of axial symmetry and search for an
upper limit of ${\cal K}_0$ in general. Guided by a mode matching
argument~\cite{SMK,EK}, we consider a directional dipolar incident
wave. In this case the incident field stems from the emission
pattern at the GRS of a virtual dipole parallel to the $x$-axis and
placed at the origin ~\cite{SD}. Following Eq.~(\ref{eq-K}), we
obtain~\cite{Sup} \beq
 {\cal K}_0= \frac 12 \left( 4 - 3 \cos \alpha - \cos^3 \alpha \right)~.
\label{eq-A-K} \eeq We remark that $\cal A$ deduced from
Eqs.~(\ref{eq-K})-(\ref{eq-A-K}) is equivalent to the corresponding
expression for the normalized energy density in Ref.~\cite{ST,She}.
At $\alpha = \pi/2$, ${\cal A}$ reaches its minimum value of ${\cal
A} = \sigma_0/2$ and ${\cal K}_0$ its ultimate  maximum
value of 2, respectively.  This value is consistent with the limit
${W}_{\rm inc}({\rm O})/P_{\rm inc} \le k^2/(3 \pi c)$ given by
Bassett for the sum ${W}_{\rm inc}$ of the time-averaged electric
and magnetic energy densities at the focal spot \cite{Bas}. As a
last case study we consider the interaction of an oscillating dipole
oriented along the $z$-axis with a radially polarized dipolar
incident field obtained from the radiation of a virtual dipole
oriented along the $z$-axis and located at O~\cite{SL}. Here too, we
find that ${\cal K}_0$ reaches the maximum value of 2 at
$\alpha=\pi/2$.

\begin{figure}
\includegraphics[width=0.95\columnwidth]{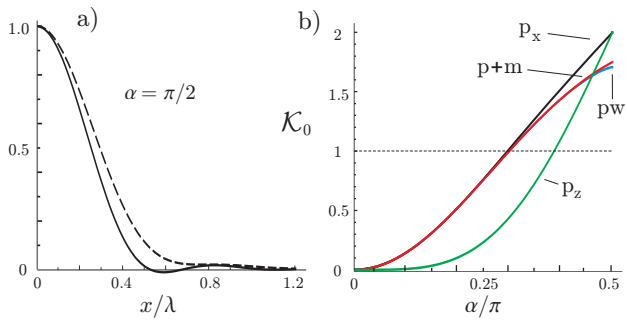}
\caption{ a) The $z$-component of the Poynting vector $S_z$ (full
curve) and the electric energy density ${W}_{\rm inc}^{\rm el}$
(dashed curve), both along the positive $x$-axis in the focal plane
and normalized to their respective values at $x=0$. b) Scattering
ratio ${\cal K}_0$ as a function of $\alpha$ for several focused waves. $\rm p_z$ denotes
the dipole wave with the electric dipole along the $z$-axis, pw
signifies the focused plane wave, $\rm p+m$ shows the
combined electric and magnetic dipole fields, and $\rm p_x$ notes
the wave of an electric dipole oriented along the $x$-axis.
${\cal K}_0=2$ is reached at $\alpha=\pi/2$ for $\rm p_x$ and $\rm
p_z$. The horizontal dashed line marks the regime of ${\cal K}_0\ge 1$.}
\end{figure}

Fig.~2b displays  ${\cal K}_0$ as a function of
$\alpha$ for various illuminations considered above. In all cases, $
{\cal K}_0 \gtrsim 1 $ is met for realistic numerical apertures. We
are, thus, facing the paradoxical seeming situation that the power
emitted by the oscillator may be larger than the incident power.
However, this finding does not violate the law of power conservation
because there is destructive interference in the forward direction.
We analyze this interference by determining now the incident
and scattered fields at the GRS for $z>0$. A particularly insightful
approach is to expand an arbitrary excitation field in terms of
vectorial multipoles \cite{BH,ST1,MSA}. All multipoles
become zero at the origin except the electric
dipole mode, which for a transverse system reads \cite{BH}
\beq
  {\bf N}_{{\rm e}11} = \left\{ \begin{array}{ll}
     \frac 23 {\bf \hat e}_x ,& {\bf r} = {\rm O}\\
    \left( \cos \vartheta \cos \varphi \, {\bf \hat e}_\vartheta
   - \sin \varphi \, {\bf \hat e}_\varphi \right)\frac{ \D e^{ i (kr-\pi/2)}}{kr}
  ,& kr \gg 1 .
\end{array} \right. \label{eq-Ne}
\eeq
We note that here the field for $kr \gg 1$ is given only for
the outgoing wave.  The electric dipole-wave component $\BPsi$ of the
excitation field can be written as
\beq
   \BPsi({\bf r}) = \frac{ E_{\rm inc}({\rm O})}{|{\bf N}_{{\rm e}11}({\rm O})|} \,
     {\bf N}_{{\rm e}11}({\bf r})~,
\eeq
where $E_{\rm inc}({\rm O})$ is taken from the Debye diffraction approach~\cite{Sup}. The field
scattered by the oscillator also forms a dipole wave \cite{Jac}
\beq
   {\bf E}_{\rm sca}({\bf  r}) = - \frac{3\Gam  E_{\rm inc}({\rm O}) }{ 2 (2 \Delta + i \Gam)}
    ~\frac {e^{ik r}}{kr} ~\left[ {\bf \hat e}_x - ( {\bf \hat e}_x \cdot {\bf \hat r} \right )
      {\bf \hat r}]~,
\label{eq-Escat} \eeq where ${\bf \hat r}$ is the unit vector along
$\bf r$, and the polarization of ${\bf E_{\rm inc}}({\rm O})$ is
along the $x$-axis. At resonance one finds ${\bf E}_{\rm sca} = -
\BPsi$ for $kr \gg 1, z>0$. Therefore, the dipole wave component of
the excitation field is completely reflected just as in the
reflection of a collimated beam from a perfect metal. The $\pi$
phase shift of ${\bf E}_{\rm sca}$ with respect to ${\bf E}_{\rm
inc}$ results from the sum of two effects. First, the comparison of
Eqs. (\ref{eq-Ne}) and  (\ref{eq-Escat}) reveals a relative Gouy
phase shift of $-\pi/2$ \cite{BW}. Second, the denominator of the
Lorentzian term in Eq. (\ref{eq-Escat}) yields a phase shift of
$\pi/2$ for $\Delta=0$, as is common for an oscillator driven at
resonance.

\begin{figure}
\centerline{\includegraphics[width=1.\columnwidth]{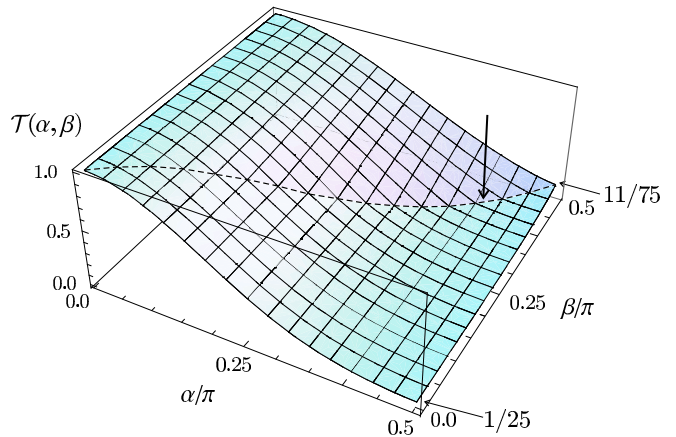}}
\caption{ Transmittance $ {\cal T}$ of a focused plane wave as a
function of the angles $\alpha$ and $\beta$ as defined in Fig.~1a.
The dashed curve indicates the edge at the geometrical shadow
boundary $\alpha=\beta$ and the vertical arrow indicates the
location $\alpha=\beta \simeq 0.43 \pi$ of the minimum value equal
to $10\%$. $ {\cal T}$-values for the cases $\alpha= \pi/2,
\beta\rightarrow 0$ and $\alpha=\beta= \pi/2$ are also noted. }
\end{figure}

This approach allows for an easy calculation of the transmittance
$\cal T$ as a function of the angles $\alpha$ and $\beta$.
For a focused incident plane wave, we find \cite{Sup}
\beq
\begin{array}{l}
 \D {\cal T}(\alpha,\beta) =
      1 +  \frac {3 {\cal I}_0(\alpha)}{2 \sin^2 \gamma}  \left[
      X(\beta) {\cal I}_0(\alpha)  -  {\cal I}_0(\gamma) \right] ~, \\
  X(\beta) = \frac 1 8 \left(4 - 3 \cos \beta - \cos^3 \beta \right)~,~
  \gamma=\min\{ \alpha, \beta \}~, \\
   {\cal I}_0(\xi) = \frac {16} {15}\left[ 1  - \frac 18  ( 5 +
   3 \cos \xi )\cos^{3/2} \xi \right] ,~~ \xi \in \{\alpha, \beta \} ~,\end{array}
\label{eq-T}
\eeq
where $\Om_\alpha$ and $\Om_\beta$ are the
incident and collection solid angles, respectively. The numerical
data in Fig.~3 display a rapid decrease of $\cal T$ with increasing
$\alpha$, while the dependence on $\beta$ is less pronounced. Of
particular experimental relevance is the geometrical shadow boundary
where $\beta=\alpha$, i.e. one collects all the incident light.
Along this line the transmittance experiences a minimum of ${\cal T}
\simeq 0.1$ at $\alpha \simeq 0.43 \, \pi$. The transmittance is plotted as a
function of detuning in Fig.~1b for
$\alpha=\beta = \pi/3$.  We point out that more complicated expressions are
expected if the dipole is displaced from the focal spot.
Particularly, the phase fronts of the scattered and excitation
fields no longer match at the GRS. The different phase fronts can give
rise to dispersive shapes of the transmission spectra depending on
the position of the detector at the GRS.

In this work, we have shown that a classical point-like oscillating
dipole can undergo a strong coupling with a confined incident
beam, reaching a 100~\% efficiency when the illumination consists of a
directional dipolar field. In fact, in the limit of weak excitation many
essential features of light-matter interaction are shared by the
quantum electrodynamic and classical formalisms
alike~\cite{Hei}. A central underlying reason for this
phenomenon is that both treatments use the same spatial description
of the electromagnetic field. To this end, our classical results can
be readily extended to the interaction of light with a two-level
system. The scattering cross section of a quantum mechanical TLS is
known to be~\cite{CDG}
\beq
    \sigma_{\rm TLS} = \sigma_0  \,
    \frac{ \Gam^2}{  4 \Delta^2 + \Gam^2 + 2 \V^2 } ~,
\label{eq-sigmaTLS}
\eeq
where $\sigma_0$ is the same quantity as in
Eq. (\ref{eq-sigma}), $\Gam$ stands for the spontaneous emission
rate of the upper level, $\V = -{\bf d}_{12} \cdot {\bf E_{\rm inc}}({\rm
O})/\hbar$ is the Rabi frequency, and ${\bf d}_{12}$ denotes the
vectorial transition dipole moment. In a semiclassical
treatment, the coherently scattered field by an atom is~\cite{CDG}
\beq
   {\bf E}_{\rm sca}^{\rm coh}
  =  \frac{ -3 \Gam \left( \Delta - i \Gam/2\right) E_{\rm inc}({\rm O}) }
     {4 \Delta^2 + \Gam^2  + 2 \V^2  }
     \frac {e^{i k r}} {kr}
\left[ {\bf \hat e}_x - \left( {\bf \hat e_{\it x} \cdot  \hat r}
\right){\bf \hat  r }  \right].~~ \label{eq-Ecoh}
\eeq
At weak excitation, $|\V| \ll \Gam$,  Eqs. (\ref{eq-sigmaTLS}) and
(\ref{eq-Ecoh}) become equivalent to Eqs.~(\ref{eq-sigma}) and
(\ref{eq-Escat}). Therefore, the results obtained for the classical
oscillator also hold for a TLS. We thus conclude that a directional dipole
wave can be perfectly reflected from a TLS under weak excitation. However, in
the saturation regime, $|\V| \gtrsim \Gam$, $\sigma_{\rm TLS}$
decreases with increasing excitation.

Considering a quantized
field, we are led to conclude that a few or even single photon
pulses can be fully reflected by a single TLS if the coherence time
of the photon is sufficiently long compared to the excited state
lifetime~\cite{DHR}. The modal formalism developed in this Letter
can be extended in the context of QED to analyze such phenomena and
will be the subject of a future study. Furthermore, it would be
interesting to investigate the photon auto-correlation function
since photon bunching or antibunching are generally expected when
there is destructive or constructive interference, respectively
\cite{EK}.

In conclusion, we have shown that a single point-like oscillating
dipole can fully reflect an incident light field. For the
experimentally important case of a focused plane wave we have found
that the transmission can be attenuated by up to 85~\%. Our findings
readily hold for the whole electromagnetic spectrum and we expect
interesting applications in the detection and spectroscopy of
subwavelength objects in the infrared to radiowave domains. In the
optical range, we anticipate that a strong coupling between a single
photon and a single quantum system can be realized in a directional
focal system without the need for high finesse cavities. Such an
arrangement would open new doors for quantum information processing
using photons as information carriers.

We thank I. Gerhardt, S. G\"otzinger, J. Hwang, and G. Wrigge for
stimulating discussions. This work was supported by the Swiss
National Foundation (SNF) and ETH Zurich.

\end{document}